\documentclass[%
 reprint,
 amsmath,amssymb,
 aps,
]{revtex4-2}
\usepackage[dvipsnames]{xcolor}
\usepackage{graphicx}
\usepackage[dvipsnames]{xcolor}
\usepackage{ragged2e}
\usepackage{graphicx}
\usepackage{dcolumn}
\usepackage{bm}

\emergencystretch=\maxdimen
\begin{document}

\preprint{APS/123-QED}

\title{Topological Optical Parametric Oscillation}

\author{Arkadev Roy}
\author{Midya Parto}
\author{Rajveer Nehra}
\author{Christian Leefmans}

 \author{Alireza Marandi$^*$}
\affiliation{%
Department of Electrical Engineering, California Institute of Technology, Pasadena, California 91125, USA}%

\affiliation{$^*$marandi@caltech.edu}

\begin{abstract}
Topological insulators possess protected boundary states which are robust against disorders and have immense implications in both fermionic and bosonic systems. Harnessing these topological effects in non-equilibrium scenarios is highly desirable and has led to the development of topological lasers. The topologically protected boundary states usually lie within the bulk bandgap, and selectively exciting them without inducing instability in the bulk modes of bosonic systems is challenging. Here, we consider topological parametrically driven nonlinear resonator arrays that possess complex eigenvalues only in the edge modes in spite of the uniform pumping. We show parametric oscillation occurs in the topological boundary modes of one and two dimensional systems as well as in the corner modes of a higher order topological insulator system. Furthermore, we demonstrate squeezing dynamics below the oscillation threshold, where the quantum properties of the topological edge modes are robust against certain disorders. Our work sheds light on the dynamics of weakly nonlinear topological systems driven out-of-equilibrium and reveals their intriguing behavior in the quantum regime.  
\end{abstract}  

\maketitle

Topological invariance and its consequences on material properties originally studied in condensed matter physics has expanded its ambit and has been investigated in diverse fields \cite{ozawa2019topological, lu2014topological, hafezi2013imaging, khanikaev2017two, rechtsman2013photonic}. Of prime interest is the presence of topologically protected edge modes which inherit their robustness from the non-trivial topology of the bulk. The introduction of topology in photonics promises a number of functionalities that leverage backscatter-free unidirectional light transport of such modes\cite{hafezi2011robust}.  

Topologically protected edge states usually lie within the bulk bandgap. Topological consequences in particle conserving fermionic Hamiltonians are manifested when the Fermi energy level lies within this bandgap. Under these circumstances the near-equilibrium dynamics are dictated by the degrees of freedom associated with the boundary/edge modes of the system. On the contrary, for bosonic systems, particles tend to condense in the lowest bulk band, resulting in the system dynamics being largely unaffected by the edge states. Thus ensuring selective participation of edge modes in bosonic topological insulator requires special attention, and most experimental investigations for these systems involved directly exciting bosons into these edge modes \cite{hafezi2013imaging}. 

The topological insulator laser is a bosonic system driven out of equilibrium where lasing happens preferentially in the edge modes. These topologically protected lasers can outperform their topologically trivial counterpart in terms of slope efficiency, coherence, and robustness against disorders \cite{amelio2020theory, harari2018topological, bandres2018topological}. To suppress the bulk modes in the topological lasing, the edge modes should be unstable while the bulk bands maintain stability. One potential approach in this regard is non-Hermitian PT symmetric Hamiltonian engineering of 1D systems \cite{parto2018edge}. However, this cannot be extended in general to higher dimensions. Another approach involves selective pumping of the edge sites to realize topological lasing \cite{harari2018topological, bandres2018topological}. Nonetheless, it is highly desirable to achieve topological edge-gain effect under uniform pumping \cite{song2020p, bahari2017nonreciprocal}. 

Recently, much attention has been devoted to the rich interplay between optical nonlinearities and topology \cite{smirnova2020nonlinear, kruk2019nonlinear}. The addition of nonlinearity leads to a variety of intriguing possibilities such as nonlinearity-driven topological phase transitions \cite{maczewsky2020nonlinearity} and self-localization \cite{leykam2016edge}. Incorporating parametrically driven quadratic nonlinearity into a topological system can result in topologically protected parametric amplification and chiral inelastic transport \cite{peano2016topological1, peano2016topological2, chaudhary2021simple, bardyn2016chiral}. Owing to the particle non-conserving nature of these parametric interactions, it can cause exponentially growing bosonic occupation, which in the suitable regime of operation can lead to edge-only dynamic instability \cite{barnett2013edge, galilo2015selective}. This parametric gain being inherently instantaneous is devoid of slow carrier dynamics that can cause deleterious effects in case of topological lasers based on semiconductor gain medium \cite{longhi2018presence}. Exploiting the squeezing dynamics in these quadratic nonlinear systems also opens new avenues for leveraging topological effects in the quantum regime. Topological insulating phenomenon when applied to the quantum states can result in the topological protection of quantum statistics and quantum correlations \cite{tambasco2018quantum, rechtsman2016topological,blanco2018topological}. 

In this work,  we show topological optical parametric oscillation in a network of quadratic nonlinear resonators where the underlying linear Hamiltonian is topological. We drive the system in a parameter regime where it exhibits edge instability under uniform excitation. We demonstrate edge mode parametric oscillation in both 1D and 2D topological insulating systems as well as corner mode parametric oscillation in a higher order topological insulator. We show the presence of squeezed quantum state below the oscillation threshold which is robust against perturbations arising from symmetry-preserving disorders.  \\

\begin{figure}[!ht]
\centering
\includegraphics[width=0.5\textwidth]{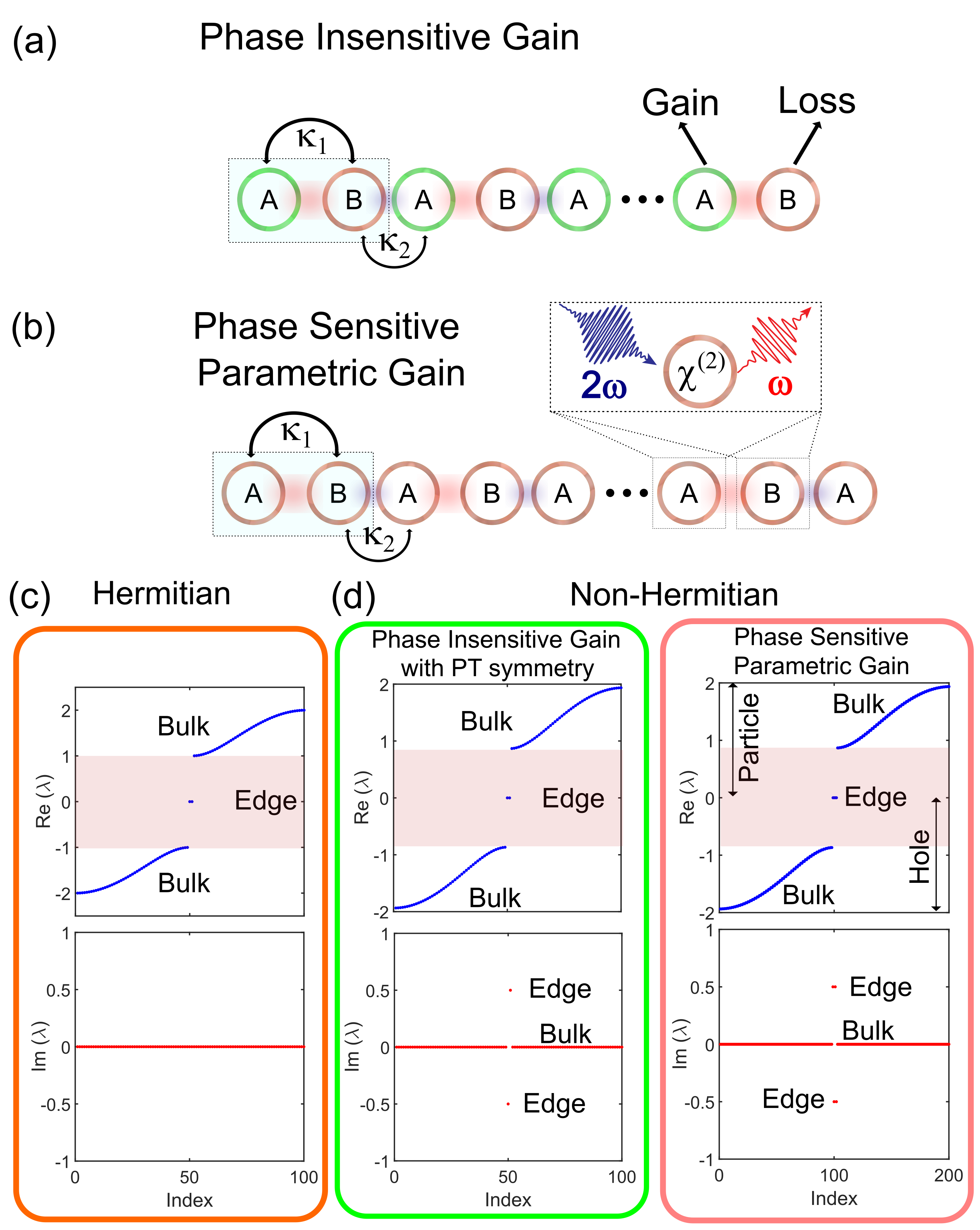}
\caption{\label{fig: schematic} \textbf{Topological edge-gain induced by parametric interaction.} a) Conventional topological edge-gain realized with alternate phase-insensitive laser gain and loss arrangement in a SSH chain of resonators. b) Similar edge-gain can be implemented with uniform parametric gain in a SSH chain of nonlinear resonators, where the quadratic nonlinearity ($\chi^{(2)}$) couples the pump ($2\omega$) and the signal ($\omega$) waves. c) Eigenvalue diagram (real part (top) and imaginary part (bottom)) for a Hermitian SSH model showing the topologically protected edge mode lying within the bandgap. Edge-only instability is not possible in this case. d) Eigenvalue diagram for a non-Hermitian setting realized either utilizing phase-insensitive gain (left) as shown in (a) or using the phase-sensitive parametric gain as shown in (b). The eigenvalue diagram depicts the edge-gain effect where the topologically protected mode has non-zero while the bulk modes have zero imaginary part respectively.  } 
\end{figure}

We aim to achieve topological edge-gain effect using parametric interactions induced by quadratic nonlinearity. To illustrate the scenario, we begin with 1D Su-Schrieffer-Heeger (SSH) model, which was originally proposed for the explanation of mobile neutral defects in polyacetylene \cite{su1979solitons}. The SSH model consists of a chain of coupled sites with intra-cell coupling ($\kappa_{1}=J(1-\epsilon)$) and inter-cell coupling ($\kappa_{2}=J(1+\epsilon)$), where $\epsilon$ denotes the asymmetry in the hopping strengths. Provided $\kappa_{1}<\kappa_{2}$, the 1D lattice hosts topological edge modes. In the Hermitian case, the edge mode lies within the bulk bandgap as shown in Fig. 1(c), and the application of uniform laser gain leads to bulk instability. To ensure edge-only lasing, non-Hermitian PT symmetric Hamiltonian can be engineered with alternate gain and loss as shown in Fig. 1(a) \cite{parto2018edge, zhao2018topological}. Figure 1(d) shows that under the application of this non-uniform phase-insensitive laser gain, the edge mode selectively experiences instability while the bulk modes remain stable ($\textrm{Im}(\lambda)=0$). Alternatively, we can exploit the unitary correspondence between non-Hermitian dynamical systems and the Heisenberg equations of motion governing parametrically driven quadratically nonlinear systems to achieve the topological edge-gain effect \cite{wang2019non, roy2021nondissipative}. The system resembles a lattice of coupled nonlinear resonators that experience uniform parametric gain as shown in Fig. 1(b). The parametric amplification/ de-amplification can replicate the dynamics of non-Hermitian PT symmetric SSH model in a non-dissipative setting as shown in Fig. 1(b) and Fig. 2(a). 

\section{Results}
\subsection{Classical mean-field regime}
We consider quadratic ($\chi^{(2)}$) nonlinear resonators with phase-matched parametric interaction between the pump at $2\omega$ and signal at $\omega$. The nonlinear part of the Hamiltonian is given by: 
\begin{equation}
 \hat{\mathcal{H}}_{NL}=\sum_{n} \frac{g}{2}(\hat{a}_{n}\hat{a}_{n}+H.c.)
\end{equation} 
where, $g$ is the strength of the parametric interaction that depends on the effective $\chi^{(2)}$ nonlinearity and the incident pump power. $\hat{a}, \hat{a}^{\dagger}$, represents the annihilation and the creation operators, respectively. The linear part of the Hamiltonian according to the SSH model is given by:
\begin{equation}
 \hat{\mathcal{H}}_{L}=J\sum_{n}\left (1+\epsilon(-1)^{n} \right) \left (\hat{a}_{n}^{\dagger}\hat{a}_{n+1}+H.c.\right)
\end{equation}
For a lattice of N sites, the full Hamiltonian can be expressed as $\hat{\mathcal{H}}=\frac{1}{2}\psi^{\dagger}H\psi$, where $\psi=(\hat{a}_{1},\dots\hat{a}_{N},\hat{a}_{1}^{\dagger},\dots\hat{a}_{N}^{\dagger})^{T}$, and $\hat{\mathcal{H}}=\hat{\mathcal{H}}_{L}+\hat{\mathcal{H}}_{NL}$. $H$ is the 2N $\times$ 2N Bogoliubov-de Gennes (BdG) Hamiltonian. The dynamics of the system is determined by the eigenvalues of the BdG equation: $\sigma_{z}H\psi_{n \pm}=\pm E_{n}\psi_{n \pm}$, where $\sigma_{z}$  is the Pauli matrix, $\psi_{n \pm}$ is a 2N dimensional eigenvector \cite{barnett2013edge}. $E_{n}$ are the eigenenergies. Because of the particle-hole symmetry of the BdG Hamiltonian, the eigenvalues $\pm E_{n}$ appear in pairs. For each pair of real eigenvalues, we can identify the eigenvector as being a particle with a positive norm $\psi_{n +}^{\dagger}\sigma_{z}\psi_{n +} > 0$, and the other being a hole with a negative norm $\psi_{n -}^{\dagger}\sigma_{z}\psi_{n -} < 0$ \cite{galilo2015selective}.
 \begin{figure*}[!ht]
\centering
\includegraphics[width=0.8\textwidth]{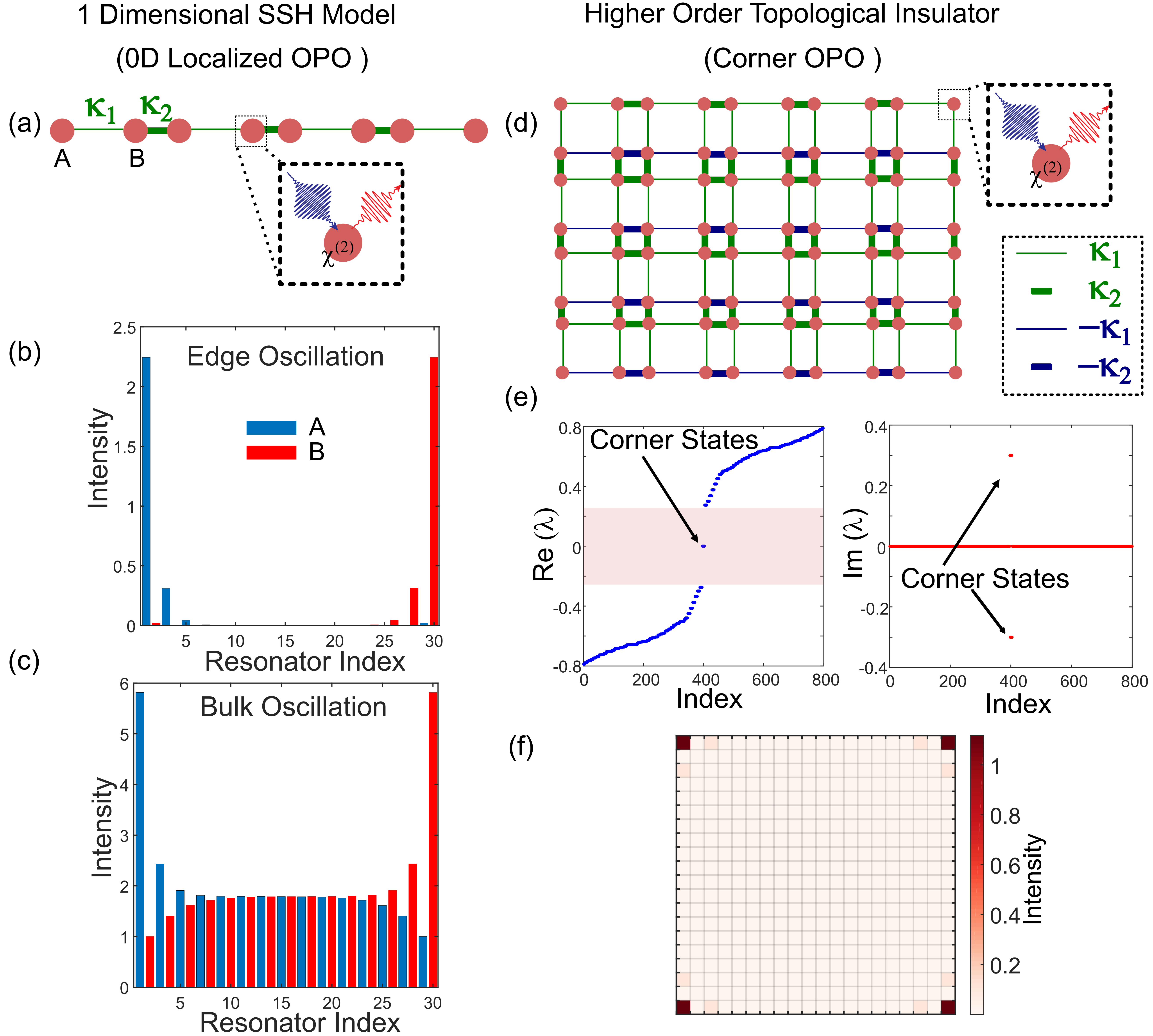}
\caption{\label{fig: Fig2} \textbf{Non-propagating topological parametric oscillation}. a) Schematic of 1D SSH model of quadratic nonlinear resonators hosting 0D localized edge modes, where the intra-cell coupling ($\kappa_{1}$, thin lines) is weaker than the inter-cell coupling ($\kappa_{2}$, thick lines). b) parametric oscillation when only the edge modes experience instability, while the bulk modes are stable. c) oscillation when the parametric gain also induces instability in the bulk modes. d) Schematic of 2D lattice of quadratic nonlinear resonators hosting corner modes. Thick and thin lines indicate strong and weak coupling respectively, while green and blue colour represents positive and negative couplings. e) Eigenvalue diagram (the real part (left) and the imaginary part (right)) showing the corner states lying within the bandgap with non-zero imaginary part while the bulk modes are stable. f) parametric oscillation in the corner modes of the 2D lattice.     }
\end{figure*}

If $g<2J\epsilon$, then we can ensure that the parametric gain will only induce instability in the edge modes, while the bulk will remain stable (see Supplementary Section 1). The eigenvalue distribution in such a scenario is shown in Fig. 1(d). When the parametric gain experienced by the edge modes exceed the net loss, there will be growth of photon number in the edge eigenmode. However, the onset of gain saturation prevents the exponential growth and the parametric oscillation reaches the steady state (see Supplementary Section 4). The steady state intensity distribution of the oscillating supermode is shown in Fig. 2(b). The intensity distribution bears the characteristics of the edge mode as evident from the alternate site occupation. However, if $g>2J\epsilon$, the bulk modes will also become dynamically unstable (see Supplementary Section 1). The steady state intensity distribution in this scenario is shown in Fig. 2(c). \\

\begin{figure}[!ht]
\centering
\includegraphics[width=0.5\textwidth]{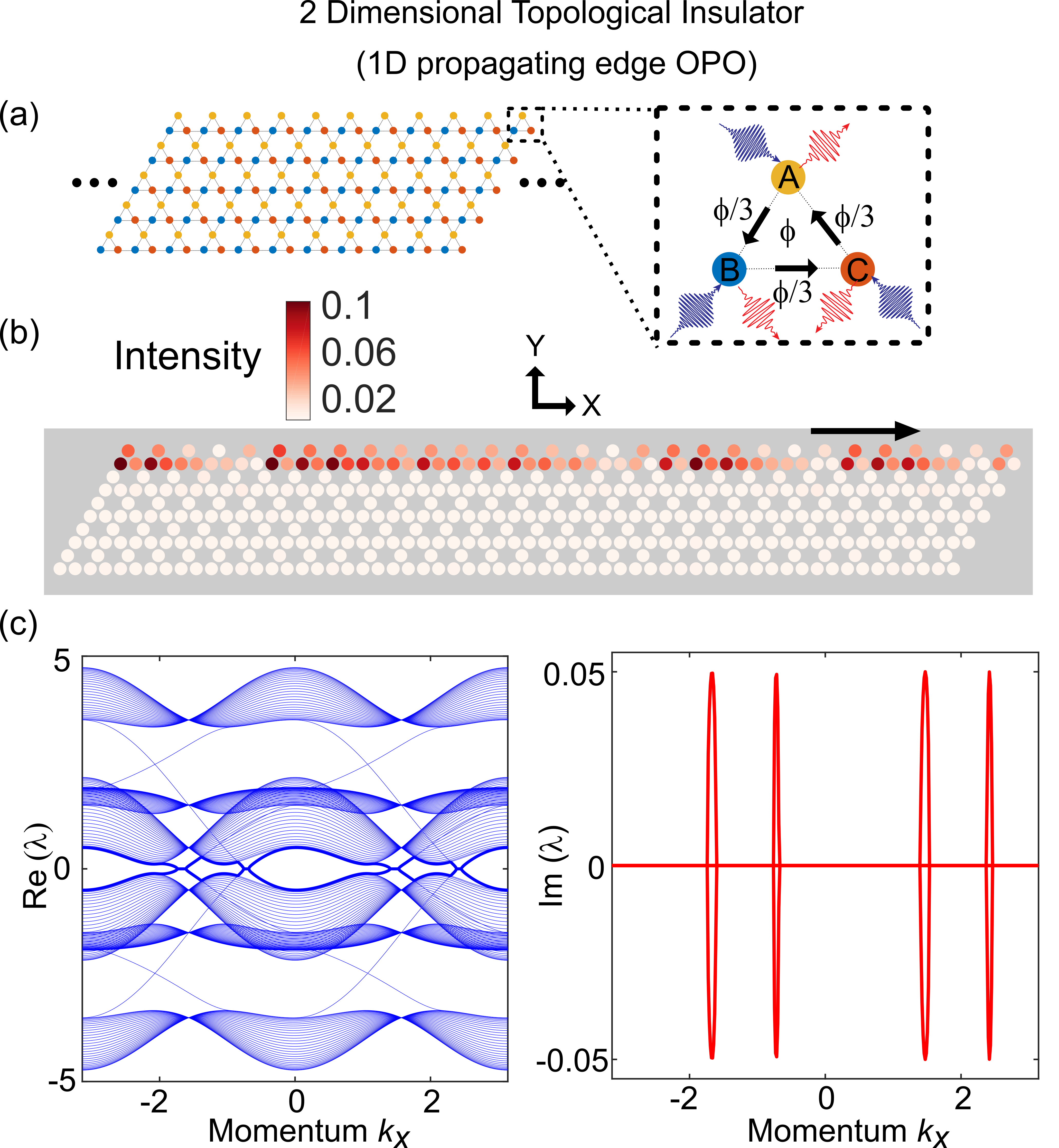}
\justifying
\caption{\label{fig: Fig3}  \textbf{Chiral topological parametric oscillation.}  a) Schematic of an anomalous quantum hall effect topological insulator arranged in the form of Kagome lattice. The zoomed in view shows the nonlinear resonators with parametric gain and the gauge flux ($\phi$). b) parametric oscillation in the chiral edge state of a one dimensional strip of the 2D Kagome lattice (finite in y, and periodic boundary condition along x). c) Band diagram (the real part (left) and the imaginary part (right)) showing the edge modes experiencing gain.   }
\end{figure} 

 Next we consider a 2D lattice of quadratic nonlinear resonators with quantized quadrupole moment \cite{mittal2019photonic1}. The  higher order topological insulator can host localized corner states. The schematic of the 2D lattice featuring couplings with different hopping strengths and phases is shown in Fig. 2(d). These couplings can be realized in a photonic platform using auxiliary non-resonant coupling cavities \cite{hafezi2013imaging}. Under uniform parametric pumping, it is possible to induce selective instability in the corner states, while the bulk modes remain stable. The eigenvalue distribution in such a scenario is shown in Fig. 2(e). When the parametric gain is above the oscillation threshold, the system can undergo parametric oscillation in the corner modes as shown by the lattice intensity distribution in Fig. 2(f). \\ 
 
So far we have considered parametric oscillation in localized topological edge states. Now we explore parametric oscillation in chiral propagating edge states. We consider an infinite strip of nonlinear resonators (finite in Y, infinite/ periodic boundary condition in X) arranged in the form of Kagome lattice as shown in Fig. 3(a). A gauge flux of $\phi$ is enclosed in each triangular plaquette. The system corresponds to an anomalous quantum Hall topological insulator \cite{mittal2019photonic2}. The linear part of the Hamiltonian is given by:

\begin{equation}
 \hat{\mathcal{H}}_{L}=J\sum_{n} \Delta \hat{a}_{n}^{\dagger}\hat{a}_{n} + \sum_{<n,n^{'}>}\kappa_{n,n^{'}} \hat{a}_{n}^{\dagger}\hat{a}_{n^{'}}
\end{equation}

where n=$(n_{x},n_{y},s)$ is the vector site index, with $n_{x}, n_{y}$ indicating the position of the unit cell in the 2D lattice, while the index $s \in (A,B,C)$ indicates the component of the sublattice. $<n,n_{'}>$ denotes the sum of the contributions over nearest neighbours. $\Delta$ is the on-site detuning. The hopping term $\kappa_{n,n^{'}}$ is given by $\kappa_{n,n^{'}}=J\textrm{e}^{\textrm{i}\phi_{s,s^{'}}}$ where $\phi_{AB}=\phi_{BC}=\phi_{CA}=\frac{\phi}{3}$. The nonlinear part of the Hamiltonian ($\hat{\mathcal{H}}_{NL}$) is same as before. This nonlinear topological insulator is characterized by the symplectic Chern numbers \cite{chaudhary2021simple}. With appropriate values of the detuning ($\Delta$) and the gauge flux ($\phi$) the particle and hole bands cross each other (see Supplementary Section 2). Thus, it creates a bandgap where the chiral edge states cross the zero energy. In the presence of non-zero parametric interactions the chiral edge states can develop instability. This is displayed in the band diagram shown in Fig. 3(c). The parametric oscillation obtained in this case is confined along the edge as depicted in the lattice intensity distribution in Fig. 3(b). The chiral nature can be observed in the dynamic evolution of the parametric oscillation (see Supplementary Section 2).  \\

\begin{figure}[!ht]
\centering
\includegraphics[width=0.5\textwidth]{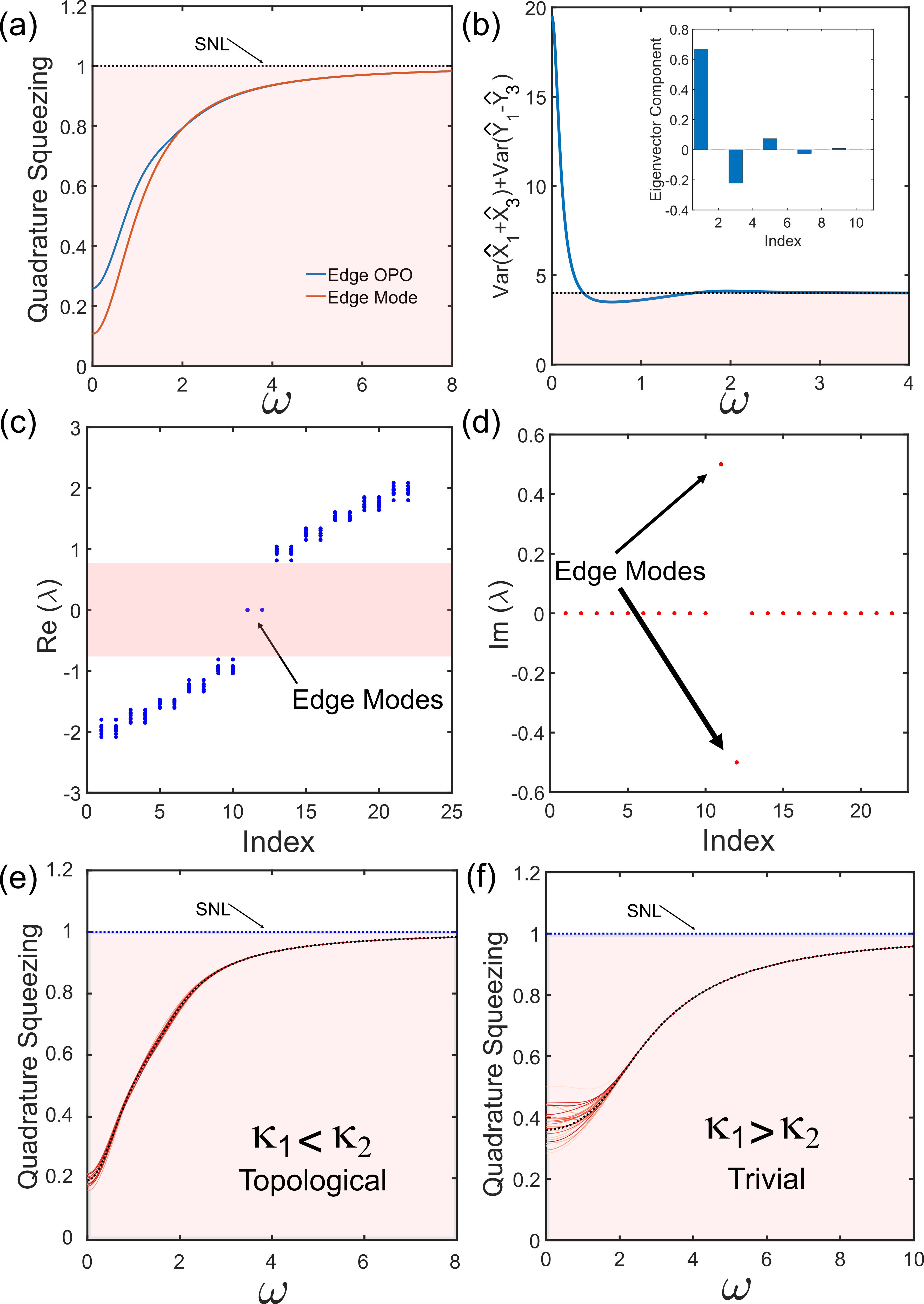}
\justifying
\caption{\label{fig: Fig4}  \textbf{Topological protection in the quantum regime.}  a) Quadrature squeezing observed in the topological edge mode of the 1D SSH model of the quadratic nonlinear resonators when the parametric gain is below the oscillation threshold. The edge super mode spans over multiple resonators (inset of (b) showing the edge eigenvector), while the edge OPO refers to the boundary resonator only. SNL represents the shot noise limit. b) Entanglement between the first and the third resonator of the edge mode as evident from the Duan criterion, where the sum of the variances is less than 4. Eigenvalue diagram c) real part and d) imaginary part showing the robustness of the edge mode to the coupling disorders. Multiple eigenvalue diagrams corresponding to different coupling disorder realizations are overlaid together. e) Protection of the quadrature squeezing in the presence of coupling disorders in the topological case ($\kappa_{1}<\kappa_{2}$). f) The trivial case ($\kappa_{1}> \kappa_{2}$) is more susceptible to squeezing degradation in the presence of coupling disorders. Multiple squeezing spectrum calculated in the fixed quadrature under different disorder realizations are overlaid together. The black dotted line corresponds to the squeezing spectrum in the absence of disorder.    }
\end{figure}  
\subsection{Quantum regime}
Below the oscillation threshold, optical parametric oscillators (OPOs) can display quadrature squeezing, where the noise in one quadrature is squeezed below the shot noise limit, and the excess noise is accumulated in the orthogonal quadrature \cite{wu1986generation}. Squeezed quantum states are an important resource in sensing \cite{aasi2013enhanced}. However, preserving the squeezing in the quadrature of interest is challenging. The occurrence of detuning or any other disorder can rotate the optimum squeezing quadrature and the squeezing in the original quadrature can be degraded due to the mixing of the anti-squeezed component. Here, we investigate the behavior of the topological quantum squeezed state in the presence of disorders. We consider the 1D SSH model (Fig. 2(a)), whose dynamics is modelled using the Heisenberg Langevin equations \cite{chembo2016quantum}. The quadrature squeezing spectrum of the topological edge mode is shown in Fig. 4(a). The edge mode is mostly confined in the boundary resonator. Thus the squeezing obtained in the edge/boundary OPO is close to that contained in the edge mode. Thus by accessing the boundary OPO, we can harness the benefits of the topological protection in the quantum regime. The edge eigenvector spans over multiple resonators, with occupations in only one of the sublattice, and alternate in sign ($\pi$ staggered) as shown in the inset of Fig. 4(b).  The adjacent A sites of the edge mode i.e. (resonators 1 and 3) are entangled as confirmed by the Duan inseparability criterion \cite{duan2000inseparability}. The entanglement is reflected in the sum of quadrature component variances as shown in Fig. 4(b). 

The bulk bands are affected by the presence of random disorders in the coupling strengths, while the edge mode exhibit robustness. This is shown in the variation of the eigenvalue diagram over multiple disorder realizations in Fig. 4(c,d). Consequently, the quadrature squeezing is protected in the topological case (Fig. 4(e)). In contrast, the non-topological case is more susceptible to squeezing degradation the presence of coupling disorders as shown in Fig. 4(f) (see Supplementary Section 3). The 1D SSH model cannot provide protection from detuning disorders which do not preserve the chiral symmetry (see Supplementary Section 3).  \\ \\
In summary, we have presented topological optical parametric oscillation in both localized and propagating chiral edge modes. We show that uniformly pumped lattice of nonlinear resonators can exhibit the topological edge-gain effect, where the edge mode undergoes instability while the bulk bands remain stable. Furthermore, we present the robustness of quantum states in the presence of symmetry-preserving disorder. Our proposed system consisting of lattice of OPOs can be realized in thin-film lithium niobate on insulator platforms \cite{lu2021ultralow}. Nanoscale OPOs can also be a promising route to realize the same \cite{jahani2021wavelength}. Alternatively, one can exploit synthetic dimensions to construct a lattice of OPOs in time/ frequency domain \cite{leefmans2021topological, dutt2020single}. Future work includes investigating the interplay of topology and nonlinearity in driven-dissipative non-equilibrium systems and its impact on emergent phenomena such as phase transitions \cite{roy2021spectral,mittal2021topological}. 

\section{Acknowledgments}
The authors gratefully acknowledge support from ARO Grant No. W911NF-18-1-0285, NSF Grant No. 1846273, NSF Grant No. 1918549 and NASA. The authors wish to thank NTT Research for their financial and technical support. \\

\nocite{*}
\bibliographystyle{unsrt} 
\bibliography{ref}


\end{document}